\documentclass[11pt,aps,numerical,graphicx,superscriptaddress,longbibliography,showkeys,utf8,onecolumn]{revtex4-2}

\makeatletter
\renewcommand{\p@subsection}{}
\renewcommand{\p@subsubsection}{}
\makeatother

\usepackage[margin=1.5in]{geometry}
\usepackage{tabularx}
\usepackage{amssymb}
\usepackage[utf8]{inputenc}

\usepackage{amsmath}
\usepackage{amsfonts}

\usepackage{geometry}

\usepackage{color}
\usepackage{caption}
\usepackage{natbib}
\usepackage{epstopdf}
\usepackage{graphicx}
\usepackage{subcaption}
\usepackage{todonotes}
\usepackage{tikz}
\usetikzlibrary{positioning}
\usepackage{float}
\usepackage{sistyle}
\usepackage[]{booktabs}
\usepackage[]{setspace}
\usepackage{hyperref}

\usepackage[english]{babel}
\usepackage{grffile}
\usepackage{tikz}

\begin{document}
 \renewcommand\texteuro{FIXME} 
\def\SymbReg{\textsuperscript{\textregistered}}

\title{\bfseries A novel analytical population TCP model includes cell density and volume variations: application to canine brain tumor }

\author{Stephan Radonic}
\email[Author to whom correspondence should be addressed. Electronic mail: ]{stephan.radonic@uzh.ch}
\affiliation{Department of Physics, University of Zurich, Zurich, Switzerland}
\affiliation{Division of Radiation Oncology, Small Animal Department, Vetsuisse Faculty, University of Zurich, Zurich, Switzerland}

\author{Jürgen Besserer}
\affiliation{Department of Physics, University of Zurich, Zurich, Switzerland}
\affiliation{Radiotherapy Hirslanden AG, Rain 34 , Aarau, Switzerland}

\author{Valeria Meier}
\affiliation{Division of Radiation Oncology, Small Animal Department, Vetsuisse Faculty, University of Zurich, Zurich, Switzerland}

\author{Carla Rohrer Bley}
\affiliation{Division of Radiation Oncology, Small Animal Department, Vetsuisse Faculty, University of Zurich, Zurich, Switzerland}

\author{Uwe Schneider}
\affiliation{Department of Physics, University of Zurich, Zurich, Switzerland}
\affiliation{Radiotherapy Hirslanden AG, Rain 34 , Aarau, Switzerland}

\date{\today}

\begin{abstract}
\textbf{Purpose:} TCP models based on Poisson statistics are characterizing the distribution of surviving clonogens. It enables the calculation of TCP for individuals. In order to describe clinically observed survival data of patient cohorts it is necessary to extend the Poisson TCP model. This is typically done by either incorporating variations of various model parameters, or by using an empirical logistic model. The purpose of this work is the development of an analytical population TCP model by mechanistic extension of the Possion model. \\
	 \textbf{Methods and Materials:} The frequency distribution of GTVs is used to incorporate tumor volume variations into the TCP model. Additionally the tumour cell density variation is incorporated. Both versions of the population TCP model were fitted to clinical data and compared to existing literature. \\
	\textbf{Results:} : It was shown that clinically observed brain tumour volumes of dogs undergoing radiotherapy are distributed according to an exponential distribution. The average GTV size was 3.37 cm$^3$. Fitting the population TCP model including the volume variation using the LQ and track-event model yielded $\alpha=0.36 ~Gy^{-1}$, $\beta=0.045~Gy^{-2}$, $a=0.9$, $T_D=5.0~d$ and $p = 0.36~Gy^{-1}$, $q=0.48~Gy^{-1}$, $a=0.80$, $T_D = 3.0~d$, respectively. Fitting the population TCP model including both the volume and cell density variation yields $\alpha=0.43~Gy^{-1}$, $\beta=0.0537~Gy^{-2}$, $a=2.0$, $T_D=3.0~d$, $\sigma=2.5$  and $p=0.43~ Gy^{-1}$, $q=0.55~Gy^{-1}$, $a=2.0$, $T_D=2.0~d$, $\sigma=3.0$ respectively.    \\
	\textbf{Conclusion:} Two sets of radiobiological parameters were obtained which can be used for quantifying the TCP for radiation therapy of dog brain tumors. We established a mechanistic link between the poisson statistics based individual TCP model and the logistic TCP model. This link can be used to determine the radiobiological parameters of patient specific TCP models from published fits of logistic models to cohorts of patients. \\ 
	  
\end{abstract}


\maketitle

\section{Introduction}

The concept of tumour control probability enables the quantification of the biologic radiation response of tumours. The TCP depends on the dose absorbed by the tumour and yields the likeliness of successful tumour control. The most widely established model is based on Poisson statistics characterizing the distribution of the surviving clonogenic cells \citep{Wiklund2014}. Mostly the linear-quadratic (LQ) model \citep{DEACON1984317, fowlerjf} is used for calculating cell survival. It allows to quantify TCP for distinct tumours in an individual. In the standard TCP model it is assumed that all the cells within the tumour absorb the same dose. 
In real-life patient plans treated with modern radiotherapy techniques such as IMRT and VMAT, dose distribution is always somewhat inhomogeneous, with varying degrees of inhomogeneity dependent on the techniques and planning constraints used. Hence the ability to evaluate the TCP for inhomogeneous dose distributions is paramount. To this end the tumour volume can be divided into subvolumes down to the size of a single voxel wherein the dose can be assumed to be constant.
Such a voxel-based TCP calculation method has been proposed by \citet{Webb1993}. 

The calculations require the knowledge of various radiobiological parameters. \citet{Qi2006} have fitted the individual Poisson statistics based TCP model to clinical human brain tumour patient survival cohort data. Thereby they obtained such a set of radiobiological parameters. Titting an individual TCP model to cohort data, however, is incorrect. This could explain the unrealistically low clonogenic cell number obtained in \citep{Qi2006}. Individual TCP models result in dose response curves which are too steep to match clinically observed survival data \citep{10.1007978-3-642-48681-4_71, Webb1993}. In these findings \citep{10.1007978-3-642-48681-4_71, Webb1993} particularly, this issue was tackled by assuming that the radio sensitivity is normally distributed among a patient population. The variation over the $\alpha$ parameters was numerically incorporated into the overall 'cohort' TCP, the fractionation was ignored. There have also been various other efforts to establish an statistical population TCP model, where variations of the radio-sensitivity, clonogenic cell density, repopulation rate and other parameters have been considered \citep{Alaswad2019, Wiklund2014, Roberts1998}.   \\
\\
In this work a novel, recently developed population TCP model \cite{Schneider2020.09.01.20185595}, has been extended and fitted to clinical dog brain tumour patient survival data from \citet{Bley2005,Schwarz2018,Thrall1999,Keyerleber2015}. The population TCP model leverages the GTV (gross tumour volume) information to incorporate the tumour volume size variation in a cohort into the Poisson statistics based individual TCP model \citep{Schneider2020.09.01.20185595}. In the context of this study it was extended to also incorporate the cell density variation. Both population TCP models were then fitted to clinical data.  To calculate the cell survival we used both the linear-quadratic and also the track event model \citep{Besserer2015}. Hypofractionation is common and established practice in the radiotherapy of tumours in small animals. The track event model is superior for calculating TCPs for high single fraction doses \citep{Besserer2015}. To our knowledge there has yet not been a comparable approach to devise a general analytic cohort TCP model.

% are currently in the process of planning a randomised clinical study on canine patients wherein a possible advantage of deliberate heterogeneous radiation dose on tumour control shall be investigated. To this end adequate models and methods are needed for determining the tumour control probability (TCP). The inhomogeneous dose distribution particularly requires voxel-based TCP calculations.

\section{Methods and Materials}
The individual TCP is modified to represent a cohort by incorporating tumour volume variations within the patient population. In a second step additionally the tumour cell density variation is included as well.\\
\\
The poisson TCP, as modelled in \citep{10.1007978-3-642-48681-4_71, Webb1993}, is determined by the number of clonogenic cells $N_S$ surviving after being exposed to a radiation dose
\begin{equation}
TCP = e^{-N_s}
\end{equation}
The number of surviving cells $N_S=N_0\cdot S$ can be calculated from the initial number $N_0$ and the cell survival function $S$, which models the dose dependence. $N_S$ can be expressed as the product of the tumour volume $V$ and the tumour cell density $\rho$.
In case of LQ Model \citep{fowlerjf} the cell survival function is given by
%and when taking in to account the fractionation by 
\begin{equation}
S^*(D,d_f,\alpha,\beta)=e^{-\alpha D - \beta d_f D}
\label{eq:frac_lq}
\end{equation}
and using the track-event model \citep{Besserer2015} the cell survival is given by:
\begin{equation}
S^*(D,d_f,q,p)=(1+q\cdot d_f)^{\frac{D}{d_f}}e^{-D\left(p+q\right)}
\label{eq:frac_pq}
\end{equation} 
where $D$ is the total radiation dose received and $d_f$ is the single fraction dose.
The LQ-model is combined with an assumed tumour repopulation factor \citep{doi:10.12590007-1285-62-735-241, Li2003}.
\begin{equation}
S(...)=S^*(...)~e^{\gamma T}
\end{equation}
Further the dependence of the survival rate on the elapse time is characterized by an exponential decrease \citep{Qi2006}.
The patient survival (TCP) after a follow-up period $\tau$ is then given by
\begin{equation}
TCP =  e^{-\rho V S e^{a\tau}}
\label{eq:ind_TCP}
\end{equation}
where $e^{\gamma T}$ accounts for the effective tumour-cell repopulation rate and $e^{a\tau}$ characterizes exponential dependence of the survival rate on the elapse time. 

The TCP model as in Eq.~\ref{eq:ind_TCP} enables the calculation of distinct tumours in individuals.

Here we derive an population TCP model by analytically incorporating variations of tumour volume sizes. Assuming an exponential frequency distribution of the GTV (gross tumour volume) size as in \citep{Schneider2020.09.01.20185595} 
\begin{equation}
f_{V_{avg}}(V)=\frac{1}{V_{avg}}\exp\left(-\frac{V}{V_{avg}}\right)
\label{eq:expvol}
\end{equation}
the population TCP is given by 
\begin{align}
TCP_{cohort} &= \int_0^\infty \exp\left(-\rho V S e^{a\tau}\right)\cdot\frac{1}{V_{avg}}\exp\left(-\frac{V}{V_{avg}}\right) dV\\
&= \frac{1}{\rho S e^{a\tau} V_{avg}+1}
\label{eq:coh_TCP}
\end{align} 
It is an interesting observation that the integration in Eq.~\ref{eq:coh_TCP} characteristically yields the logistic model, which was previously used by \citet{Okunieff1995} without further mechanistic or qualitative justification. As shown in \citep{Okunieff1995} the logistic model provides a good fit to clinical data. \\
\\
The underlying frequency distribution of present tumour volumes sizes is assumed to be exponentially distributed as in Eq.~\ref{eq:expvol}. However it is reasonable to conjecture that there is a minimal tumour volume below which tumours are unlikely to cause major symptoms and are thus unlikely to be clinically observed. Also the detectability is limited by technical and procedural constraints. Thus the exponential distribution in Eq.~\ref{eq:expvol} is modified by a switch-on term \citep{Schneider2020.09.01.20185595}
\begin{equation}
1-\exp\left(-\frac{V}{V_C}\right)
\end{equation}
where $V_C$ is a surrogate measure which characteristics the limited clinical observability of small tumour volumes. The modified distribution is then given by 
\begin{equation}
f_{V_{avg},V_C}^{obs}(V)=\frac{V_{avg}+V_C}{V_{avg}^2}\left(1-\exp\left(-\frac{V}{V_C}\right)\right)\exp\left(-\frac{V}{V_{avg}}\right)
\label{eq:vol_freq_switch}
\end{equation}
The normalization factor $\frac{V_{avg}+V_C}{V_{avg}}$ arises due to the requirement $\int\limits_{-\infty}^{\infty}f(V)dV\overset{!}{=}1$.
\\
\\
If further the model is extended by assuming an exponential tumour cell density variation inside the cohort around an average value, the tumour cell density $\rho$ is expressed as 
\begin{equation}
\rho = 10^r
\end{equation}
the cohort TCP is obtained by integrating from an $r_0 - \sigma$ to $r_0 + \sigma$, where $r_0$ is the assumed centre exponential value and $\sigma$ is the assumed variation range
\begin{align}
TCP_{cohort} &= \int_{r_0-\sigma}^{r_0+\sigma} \frac{1}{2\sigma}\frac{1}{10^r S e^{a\tau} V_{avg}+1}  dr\\
&= \frac{\log\left(S e^{a\tau} V_{avg} 10^{r_0}+10^{\sigma}\right)-\log\left(S e^{a\tau} V_{avg} 10^{r_0+\sigma}+1\right)+\sigma\log(10)}{\log(10)\sigma}
\label{eq:tcp_coh_ext}
\end{align}
In the literature \citep{Webb1993,Jin2011,Alaswad2019} an average tumour cell density of $10^7 cm^{-3}$ has been used for TCP calculations, we decided to also assume this value. \\
\\
For the fitting a least square fit was performed. The fitting parameter ranges were constrained to physically reasonable bandwidths. \\
\\

To further explore the link between the poisson TCP and the logistic TCP, the population TCP model was compared the logit fit from \citep{Okunieff1995}. Also the relations between the model parameters used by Okunieff and the model parameters used in this work are established.  

In Okunieffs logit model the TCP model is essentially characterized by two parameters. One is the $TCD_{50}$ which is the dose where half of the tumours are controlled. The second is the $slope_{50}$, which is the slope of the dose response curve at $d=TCD_{50}$ and gives an estimate of the advantage of applying an additional dose of 1 Gy to the tumour. These parameters can also be derived for the population TCP models as in Eq.~\ref{eq:coh_TCP} and Eq.~\ref{eq:tcp_coh_ext}. The derivation steps are omitted. For both population TCP models the $TCD_{50}$ can be expressed by
\begin{equation}
TCD_{50}=\frac{\log\left(\rho~V_{avg}\right)}{k}
\label{eq:tcd50}
\end{equation}
where $k=\alpha+\beta d_f$. For the volume variations based population TCP model the $slope_{50}$ is given by $slope_{50}=\frac{k}{4}$, for the extended model which also comprises the cell density variations the expression is slightly more complex 
\begin{equation}
slope_{50} = \frac{k}{2\sigma\log(10)}\frac{10^\sigma-1}{1+10^\sigma}
\label{eq:slope_50_ext}
\end{equation}
In the limit of a very small $\sigma$, $slope_{50}$ becomes the same as for the volume variation model.
\begin{equation}
\lim_{\sigma->0} \left(\frac{k}{2\sigma\log(10)}\frac{10^\sigma-1}{1+10^\sigma}\right) = \frac{k}{4}
\label{eq:limitsig}
\end{equation}

With that, radiobiological parameters of tumours can be calculated from the data published in \citet{Okunieff1995}.

\section{Results}

\setlength{\tabcolsep}{4pt}
\begin{table}[tb]

\resizebox{\columnwidth}{!}{%
\begin{tabular}{lccccccc}
\toprule
 & \multicolumn{2}{c}{Mean Dose [Gy]} & \multicolumn{2}{c}{Dose range [Gy]} & \multicolumn{3}{c}{Clinical survival after}  \\ 
& total & per fraction & total & per fraction & 12 months & 24 months & 36 months \\
\hline 
\textbf{\citet{Bley2005}} & 40.95 & 3.15 & [36.0-52.5] & [2.5-4.0] & 69 & 47 & 30 \\ 

\textbf{\citet{Keyerleber2015}} & 47.5 & 2.7 & [45.0-54.0] & [2.5-3.0] & 77 & 60 & 36 \\ 

\textbf{\citet{Thrall1999}} & 52.0 & 2.0 & [44.0-60.0] & - & 48.8 & 33 & 16 \\ 

\textbf{\citet{Schwarz2018}} & 40.0 & 4.0 & - & - & 63 & 57 & -\\

\textbf{\citet{Schwarz2018}} & 50.0 & 2.5 & - & - & 77 & 45 & -  \\

\bottomrule
\end{tabular} 
}

\caption{Clinical patient survival data, In \citep{Bley2005} and \citep{Keyerleber2015} various fractionation schemes within the denoted ranges were applied. In \citep{Schwarz2018} two distinct fractionation schemes (20 x 2.5 Gy and 10 x 4 Gy) were used exclusively, there is no follow-up data after 36 months; In \citep{Thrall1999} a fixed single fraction dose of 2 Gy was used.  }
\label{tab:clindat}
\end{table}

\subsection{Tumour volume distribution}
\label{sec:resvoldistr}
\begin{figure}[h]
\includegraphics[width=\columnwidth]{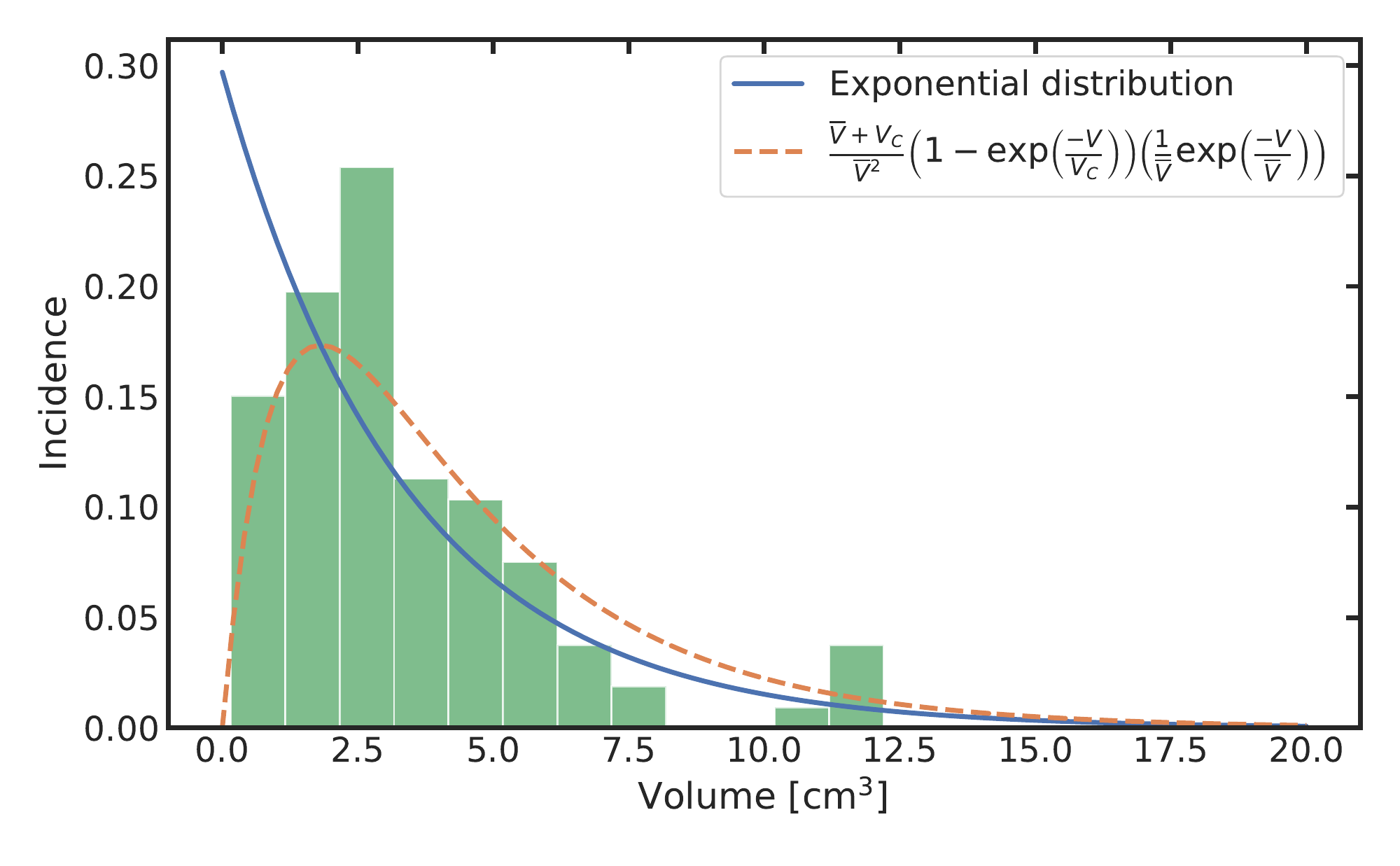} 
\caption{Histogram: clinically observed gross tumour volumes \citep{Bley2005,Schwarz2018} Full line: exponential distribution with average volume $V_{avg}$ as in Eq.~\ref{eq:expvol}, dashed line: Exponential distribution with an average volume $V_{avg}$ with an assumed limited clinical detectability governed by the parameter $V_C$ (Eq.~\ref{eq:vol_freq_switch})}
\label{fig:histo}
\end{figure}

In Fig.~\ref{fig:histo} the gross tumour volume (GTV) sizes as clinically observed by \citet{Bley2005,Schwarz2018} are plotted in a histogram. The average GTV size was fixed to the mean value of the observed clinical data $V_{avg}=\mathbf{3.37}$ cm$^3$. Further the exponential distribution (see Eq.~\ref{eq:expvol}) and a modified exponential distribution, where a limited clinical detectability, governed by the parameter $V_C$, is assumed (see Eq.~\ref{eq:vol_freq_switch}), are plotted. The value of $V_C=1.6$cm$^3$ is determined by fitting Eq.~\ref{eq:vol_freq_switch} to the clinical volume data.

\subsection{Population TCP model fitting}
\begin{figure*}
\centering
\begin{subfigure}{1.0\textwidth}
\includegraphics[height=0.45\textheight]{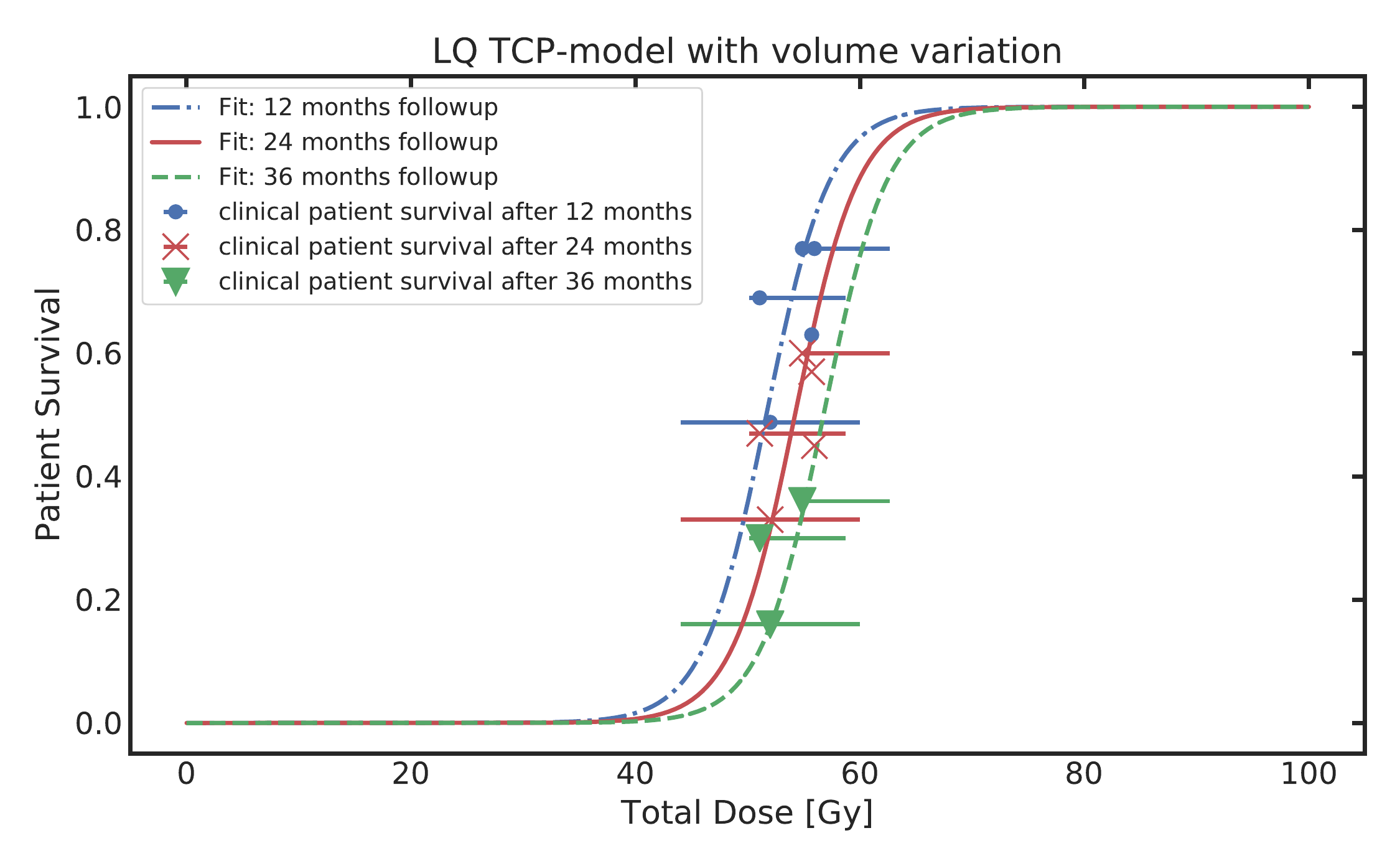} 
\subcaption{LQ cell survival model}
\label{fig:lq_vol}
\end{subfigure}
\begin{subfigure}{1.0\textwidth}
\includegraphics[height=0.45\textheight]{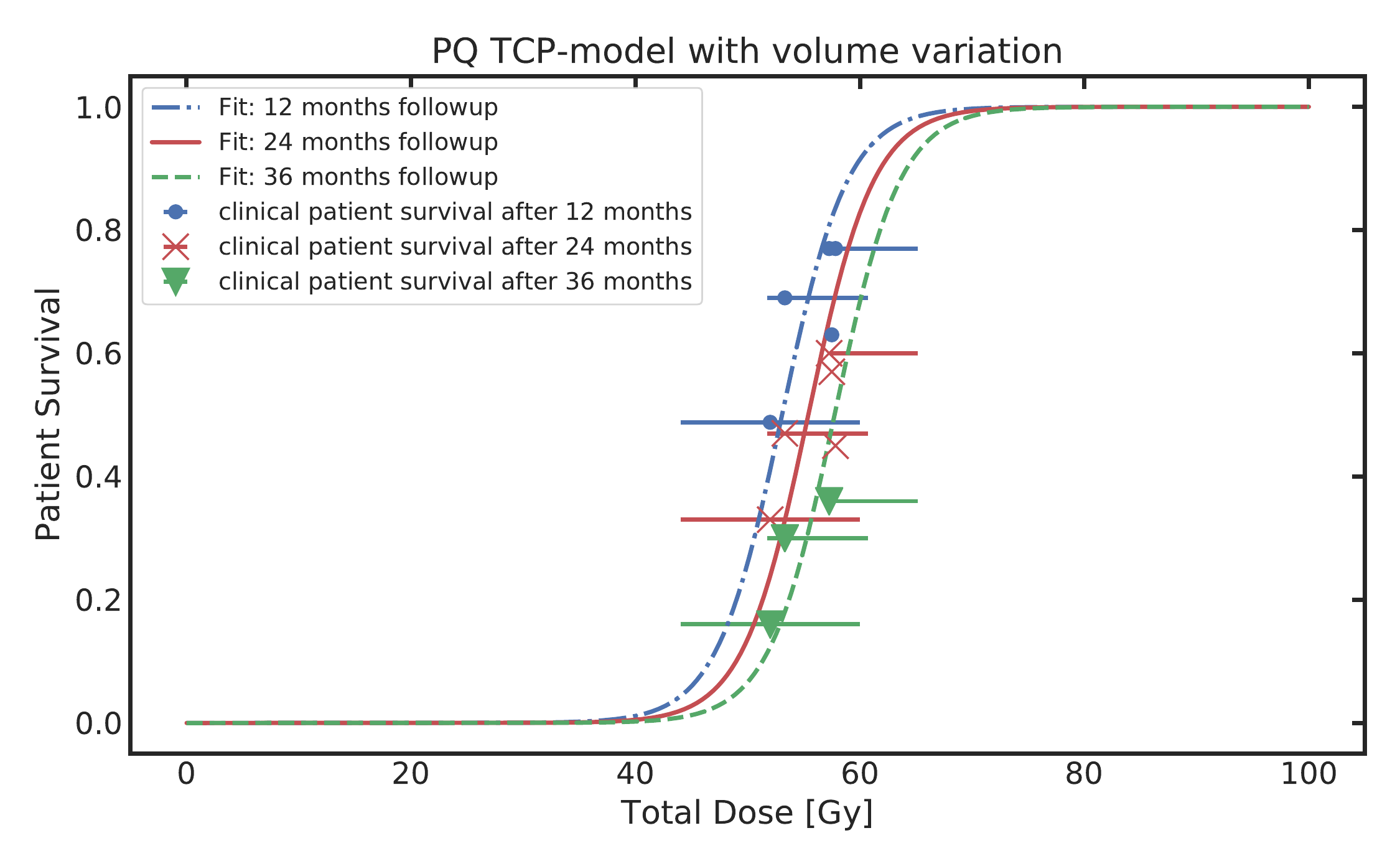} 
\subcaption{Track event cell survival model}
\label{fig:pq_vol}
\end{subfigure}
\caption{Fitting clinical patient survival with TCP population model incorporating tumour volume variations within a patient population. It should be noted that the curves in the plots for the three follow-up periods were not fitted separately but stem from a single fit to the complete data set.}
\label{fig:tcp_vol}
\end{figure*}

\begin{figure*}
\centering
\begin{subfigure}{1.0\textwidth}
\centering
\includegraphics[height=0.45\textheight]{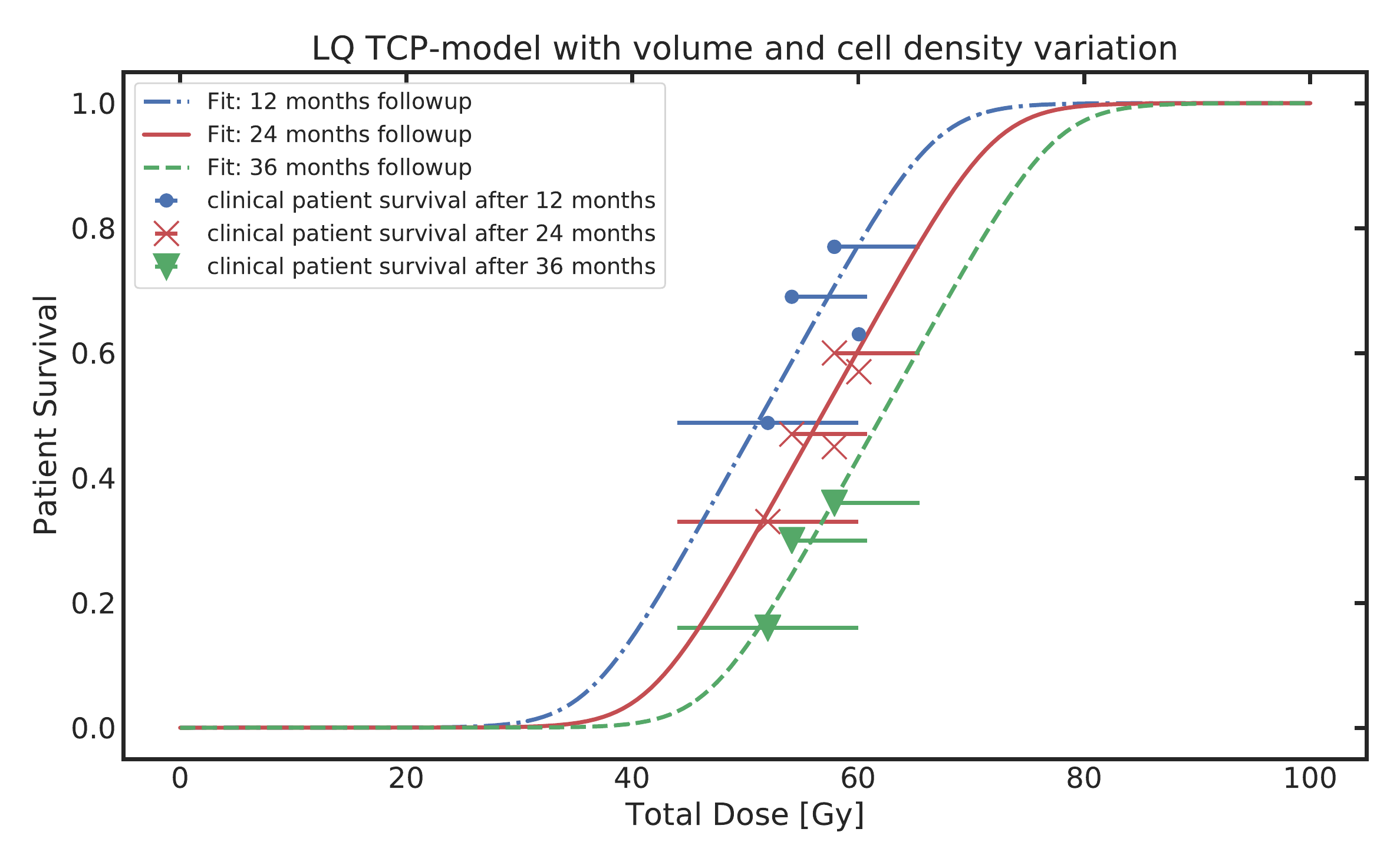} 
\subcaption{LQ cell survival model}
\label{fig:lq_12mnths}
\end{subfigure}
\begin{subfigure}{1.0\textwidth}
\includegraphics[height=0.45\textheight]{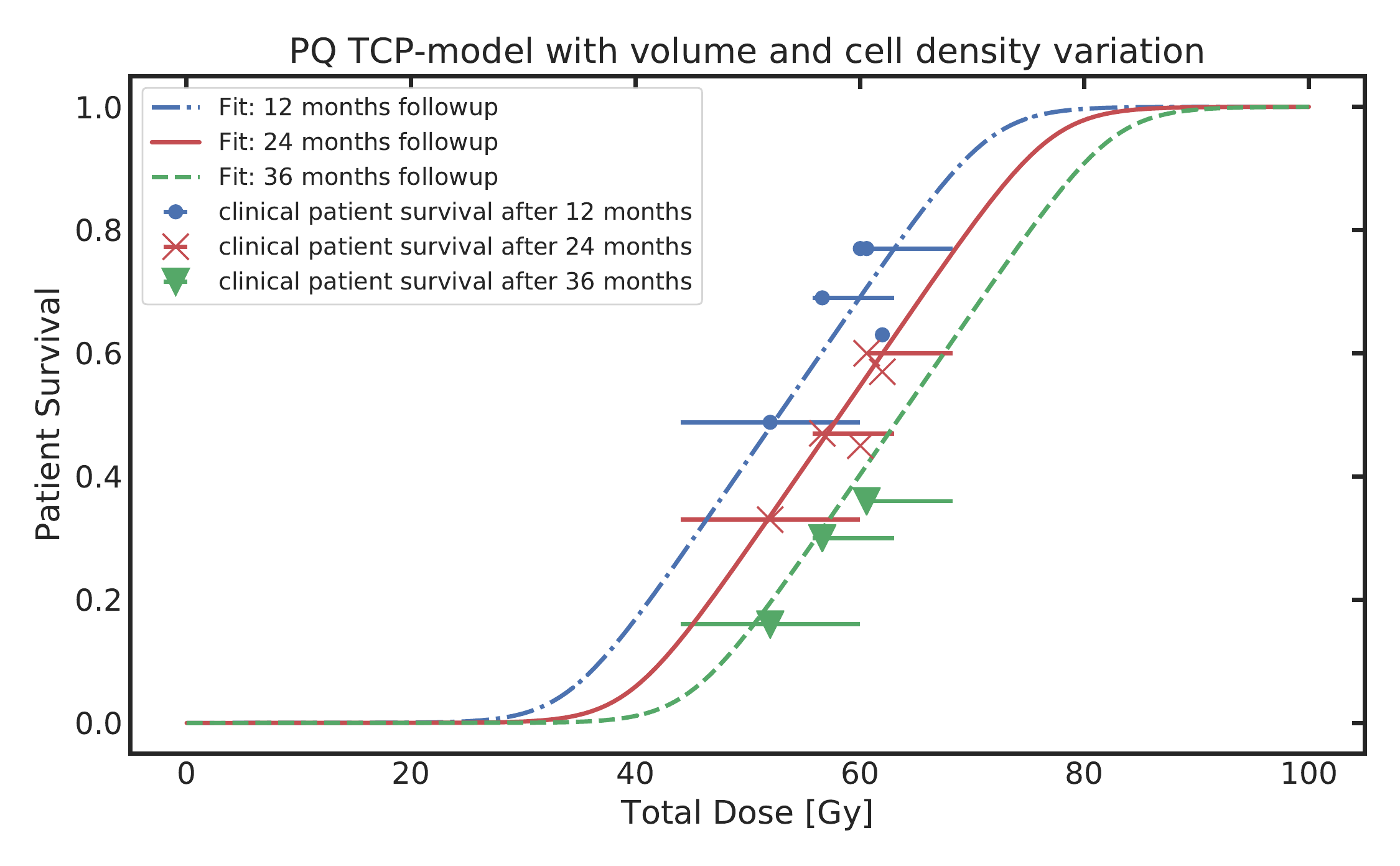} 
\subcaption{Track event cell survival model}
\label{fig:lq_24mnths}
\end{subfigure}
\caption{Fitting clinical patient survival with extended TCP population model incorporating tumour volume and cell density variations within a patient population. It should be noted that the curves in the plots for the three follow-up periods were not fitted separately but stem from a single fit to the complete data set.}
\label{fig:tcpvoldens}
\end{figure*}

The Population TCP model with tumour volume variation, as in Eq.~\ref{eq:coh_TCP} and the TCP model with both the tumour volume and tumour cell density variation, as in Eq.~\ref{eq:tcp_coh_ext} has been fitted to clinical patient survival data (Tab.~\ref{tab:clindat}) using the LQ model (Eq.~\ref{eq:frac_lq}) as well as the track event model (Eq.~\ref{eq:frac_pq}) as radiation cell survival function $S$. The average volume was set to $V_{avg}=\mathbf{3.37}$ cm$^3$ (see previous section~\ref{sec:resvoldistr}). For the model including also the cell density variation the cell density variation bandwidth $\sigma$ was a fit parameter. To decrease the number of free fitting parameters and thus increase the robustness of the fit, when using the LQ model, the $\frac{\alpha}{\beta}$ ratio was constrained to $\frac{\alpha}{\beta}=8$ Gy. In \citep{VanLeeuwen2018} a quite wide bandwidth is given for $\frac{\alpha}{\beta}$ ratios of tumours in the human central nervous system. In \citep{Pedicini2014}, where however only glioma in human patients are considered, $\frac{\alpha}{\beta}$ is listed as 8 Gy. It is established practice to use model parameters based on human data for animals. The fitting procedure of the population TCP model with the volume variation with the LQ cell survival model yields the parameter set $\alpha=0.36 ~Gy^{-1}$, $\beta=0.045~Gy^{-2}$, $a=0.9$, $T_D=5.0~d$, while the fitting with the track event model yields $p = 0.36~Gy^{-1}$, $q=0.48~Gy^{-1}$, $a=0.80$, $T_D = 3.0~d$. Fitting the population TCP model with both the volume and cell density variation yields $\alpha=0.43~Gy^{-1}$, $\beta=0.0537~Gy^{-2}$, $a=2.0$, $T_D=3.0~d$, $\sigma=2.5$ for the LQ cell survival model and $p=0.43~ Gy^{-1}$, $q=0.55~Gy^{-1}$, $a=2.0$, $T_D=2.0~d$, $\sigma=3.0$ using the track event model. An overview of the obtained parameter sets is shown in Tab.~\ref{tab:res}. In Figs.~\ref{fig:tcp_vol} and \ref{fig:tcpvoldens} the fits using both cell survival models are plotted alongside the clinical data points for follow-up periods of 12, 24 and 36 months. In the plots the original data points which had a different single fraction dose than $2$ Gy have been recalculated to $2$ Gy single fraction doses using Eqs.~\ref{eq:frac_lq} and \ref{eq:frac_pq} such that radiation cell survival matches the original prescription. It should be noted that the curves in the plots for the three follow-up periods were not fitted separately but stem from a single fit to the complete data set.

\setlength{\tabcolsep}{7pt}
\begin{table}[tb]
\resizebox{\columnwidth}{!}{%
\begin{tabularx}{\columnwidth}{lcccc}
\toprule
\textbf{parameter} & vol., LQ  & vol., PQ & vol. \& cell dens., LQ & vol. \& cell dens., PQ \\
\hline
$\alpha~[Gy^{-1}]$  & 0.36 & - & 0.43 & - \\  
$\beta~[Gy^{-2}]$ & 0.045 & - & 0.057 & - \\ 
$p~[Gy^{-1}]$ & - & 0.36 & - & 0.43 \\ 
$q~[Gy^{-1}]$ & - & 0.48 & - & 0.55 \\  
$T_d$ [d] & 5.0 & 3.0 & 3.0 & 2.0 \\ 
a & 0.9 & 0.8 & 2.0 & 2.0 \\ 
$\sigma$ & - & - & 2.5 & 3.0 \\ 
$\rho~[cm^{-3}]$  & $10^7$ & $10^7$ & $10^7$ & $10^7$ \\ 
\bottomrule
\end{tabularx} 
}
\caption{Results: radiobiological parameters from fits of the population TCP models to clinical patient survival data}
\label{tab:res}
\end{table}
% An overview of the results is shown in Table xx.

%the cell density and  $\rho=10^7 cm^{-3}$ and

\subsection{Okunieff data}

\begin{figure}[hbt]
\centering
\includegraphics[width=\columnwidth]{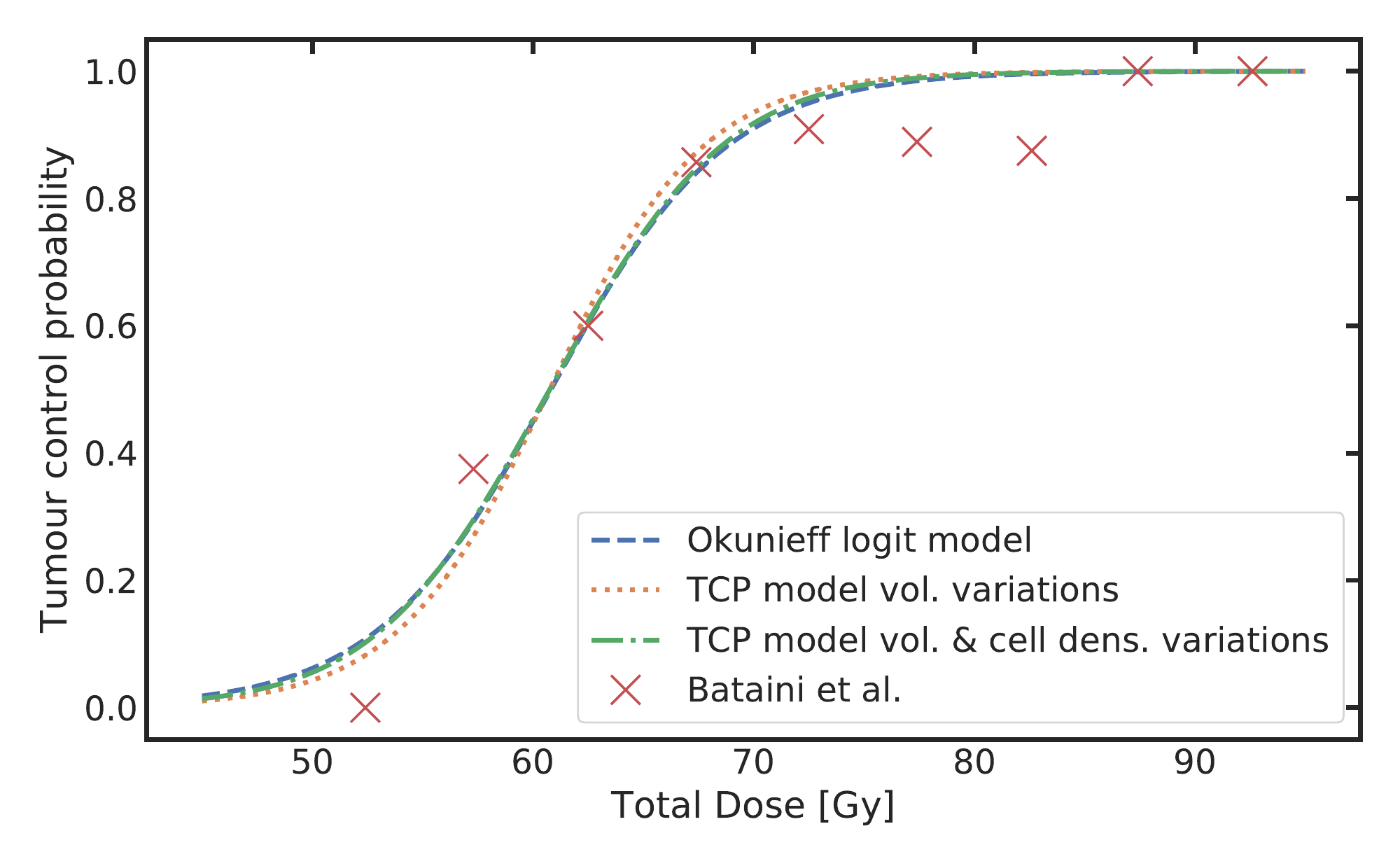} 
\caption{Comparison of Okunieffs logit fit with the population TCP models}
\label{fig:okunieffplot}
\end{figure}

In \citep{Okunieff1995}, as illustrative example, survival data \citep{BATAINI19821277} from patients treated with radiotherapy for \textit{pyriform sinus primary tumor} were plotted and fitted with the logit model. To demonstrate the equivalence of our approach, the TCP population models were also fitted to the data from \citet{BATAINI19821277}. This is shown in Fig.~\ref{fig:okunieffplot}. In \citep{Okunieff1995} $TCD_{50} = 60.8$ Gy and $slope_{50}$ of 0.063 Gy$^{-1}$ were calculated. Fitting the population TCP models with an assumed average volume of $V_{avg}=4.2$ cm$^3$ (diameter: 0.68 cm) yields $TCD_{50} = 60.75$ Gy, $slope_{50}= 0.072$ Gy$^{-1}$ for the volume variations model and $TCD_{50} = 60.75$ Gy, $slope_{50}= 0.063$ Gy$^{-1}$, $\sigma=0.6$ for the volume and cell density variations model. 

In methods and materials the relations from the population TCP models devised in this work to the logistic model used by \citet{Okunieff1995} were established. This allows the inference of population TCP model parameters ($\alpha$, $\beta$, $\sigma$) from the parameters ($TCD_{50}$, $slope_{50}$) listed in Okunieff`s work \citep{Okunieff1995}. As shown for the example of \textit{pyriform sinus primary tumour}, the parameters obtained by fitting the population TCP models to the data of \citet{BATAINI19821277} are nearly identical to Okunieffs logit fit. Hence we believe it is legitimate to calculate the radiobiological parameters from Okunieffs data. The calculated parameters for the population TCP model including tumour volume and cell density variations, using the LQ cell survival model are listed in Table~\ref{tab:okunieff}. The The $\frac{\alpha}{\beta}$ ratios required for the calculations were taken from \citep{Nahum2015}. The mean tumour cell density was fixed to $\rho=10^7$ cm$^{-3}$. Tumour sites with known diameters were selected from Table 1 from \citet{Okunieff1995}. The tumour was assumed to have a spherical shape, accordingly the tumour volume was calculated from the diameter. The calculation yields unrealistic values for the \textit{Breast tumour} with a diameter of $4-6$cm. For the \textit{Nasopharynx} with diameter of $<3$cm the calculation indicated a very small $\sigma$, thus the cell density variations was ignored in the calculation and the parameters were calculated using Eq.~\ref{eq:tcd50}, they are not consistent with Eq.~\ref{eq:limitsig}. 

\setlength{\tabcolsep}{4pt}
\begin{table}[h]
\resizebox{\columnwidth}{!}{%
\begin{tabularx}{\columnwidth}{lcccccccc}
\toprule
Tumor site       & Diameter                    & $TCD_{50}$   & $slope_{50}$  & $\alpha$    & $\beta$     & $V_{avg}$       & $\sigma$ & $\alpha / \beta$-ratio 
\\ & [cm] &  [Gy] & [\%$/$Gy] & [Gy$^{-1}$]  & [Gy$^{-1}$] & [cm$^3$] & & [Gy]   \\ \hline
Hodgkins       & \textgreater 3.0       & 15.3  & 9.09    & 1.06 & 0.084 & 14.1 & 2.9 & 12.6 \\
Hodgkins       & 0.5-3.0                & 15.31 & 9.84    & 0.87 & 0.069 & 0.5   & 2.2 & 12.6\\
Breast$^*$     & 4-6                    & 21.71 & 0.17    & -$^*$ & -$^*$  & 65.4 & -$^*$  & 6.2    \\
Breast         & \textgreater 6          & 62.59 & 1.98   & 0.25 & 0.041  & 113.1 & 3.7 & 6.2 \\
Uterine cervix & \textless 0.5 & 24.27 & 2.77    		  & 0.35   & 0.057 & 0.01   & 3.7  &  6.2 \\
Nasopharynx$^\dag$   &  \textless 3  & 50.01 & 35.06      & $0.24^\dag$ & $0.035^\dag$ & 0.5   & - & 6.9\\
Nasopharynx     & 3-6                & 55.12 & 3.31    	  & 0.28 & 0.040 & 33.5  & 2.3 & 6.9\\
Nasopharynx     & \textgreater 6     & 68.81 & 2.43       & 0.23 & 0.034 & 113.1 & 2.7 &  6.9\\
Pyriform sinus & \textless 3       & 60.76 & 6.25    	  & 0.20 & 0.029 & 0.5   & 0.2  & 6.9\\ 
Pyriform sinus & \textgreater 3   & 69.72 & 4.03          & 0.21 & 0.030 & 14.1  & 1.3 & 6.9\\
\bottomrule
\end{tabularx}
}
\caption{Radiobiological parameters $\alpha$, $\beta$ and $\sigma$ calculated from Okunieffs parameters \citep{Okunieff1995} $TCD_{50}$ and $slope_{50}$;\\ * calculation did not converge;\\ $\dag$ very small $ \sigma$ thus cell density variations ignored, parameters calculated by using Eq.~\ref{eq:tcd50}, not consistent with Eq.~\ref{eq:limitsig};}
\label{tab:okunieff}
\end{table}

%\clearpage
\section{Discussion and Conclusion}

In prior studies population TCP models were derived by numerically incorporating variations of parameters such as radio-sensitivity, clonogenic cell density, repopulation rate and others into the poisson statistics based individual TCP model \citep{10.1007978-3-642-48681-4_71, Webb1993, Alaswad2019, Wiklund2014, Roberts1998}. These studies, however, do not provide a closed form population TCP model. 
In this work we developed analytical TCP models considering variations of tumour volume and tumour cell density within a patient cohort. Then we fitted the models to clinical survival data of canine patients treated for brain tumour. It can be observed that taking into account the volume variation and even more so the density variation lead to the decrease of the TCP curve steepness. In contrast to the study of \citet{Qi2006}, the fits yielded realistic values for the clonogenic cell numbers. Most notably, to our knowledge in this work for the first time a mechanistic link has been established between the poisson statistics based individual TCP model and the logistic TCP model used to fit clinical data by e.g. \citet{Okunieff1995, Levegrun2001}. It has been shown that due to the equivalence of the population TCP models to the logistic TCP model by \citet{Okunieff1995} the radiobiological parameters ($\alpha$, $\beta$) can be calculated from the fits of the logistic model to patient data ($TD_{50}$, $slope_{50}$).  The model parameters obtained in this work can be used for assessing the prospective tumour control probability of individual radiotherapy plans. If suitable clinical data are available the model is applicable to human patients, without further constraints. \\
\\
However some limitations shall be pointed out. %A shortcoming is that the radiosensitivity variation of the tumours has not been taken into account.
For variations of parameters usually a normal distribution is assumed. For the clonogenic cell density   values between $10^4$ and $10^9$ cm$^{-3}$ are found in literature \citep{FRIEDLAND2014105,Webb1993,Jin2011,Alaswad2019, Levegrun2001}. Hence herein for the variation of the cell density we accordingly assumed an uniform variation of the exponent. This is an ad hoc assumption not attributable to literature. \\
\\ 
Fitting a multi parameter model to a limited set of data points always carries the risk of over-fitting. Thus the parameter bandwidths have been constrained to clinically reasonable ranges.
A further weakness of the fitting is that all data points lie within a relatively narrow dose range. This is  an intrinsic property of clinical data, derived from curative-intent protocols. \\
\\
Our data sample contains dogs of various sizes and breeds.  The Poisson TCP model attributes the tumour control probability to the number of surviving clonogenic cells thus a large tumour containing more cells always evaluates to a lower TCP. The absolute tumour volume is possibly correlated to the dog size or the dogs absolute brain volume. It might be fruitful to explore if the GTV size relative to the dog size is a relevant parameter to be considered in a TCP model. 

Comparing the work of \citet{Okunieff1995} to the population TCP models, average GTV sizes were estimated from the available tumour diameter information. The estimate inherently carries some uncertainty which is accordingly also comprised in the calculated parameters.

\section*{Acknowledgment}
This work was supported by the Swiss National Science Foundation (SNSF), grant number: 320030-182490; PI: Carla Rohrer Bley

\bibliographystyle{apalike}
\bibliography{literatur} 

\end{document}